\documentclass[10pt]{article}
\usepackage{graphicx}
\usepackage{amsbsy}

\title{Coupling of eigenvalues of complex matrices \\
at diabolic and exceptional points}
\author{A. A. Mailybaev, O. N. Kirillov and A. P. Seyranian}

\date{Institute of Mechanics,
Moscow State Lomonosov University, \\ Michurinskii pr. 1, 119192
Moscow, Russia \\ E-mail:
mailybaev\,\{kirillov,\,seyran\}@imec.msu.ru}

\begin{document}
\maketitle

\begin{abstract}
The paper presents a general theory of coupling of eigenvalues of
complex matrices of arbitrary dimension depending on real
parameters. The cases of weak and strong coupling  are
distinguished and their geometric interpretation in two and
three-dimensional spaces is given. General asymptotic formulae for
eigenvalue surfaces near diabolic and exceptional points are
presented demonstrating crossing and avoided crossing scenarios.
Two physical examples illustrate effectiveness and accuracy of the
presented theory.
\end{abstract}


\section{Introduction}

Behavior of eigenvalues of matrices dependent on parameters is a
problem of general interest having many important applications in
natural and engineering sciences. Probably, \cite{Hamilton1} was
the first who revealed an interesting physical effect associated
with coincident eigenvalues known as conical refraction, see
also~\cite{Berry1999}. In modern physics, e.g. quantum mechanics,
crystal optics, physical chemistry, acoustics and mechanics,
singular points of matrix spectra associated with specific effects
attract great interest of researchers since the papers
\cite{Neumann, Herring, Teller}. These are the points where
matrices possess multiple eigenvalues. In applications the case of
double eigenvalues is the most important. With a change of
parameters coupling and decoupling of eigenvalues with crossing
and avoided crossing scenario occur. The crossing of eigenvalue
surfaces (energy levels) is connected with the topic of
geometrical phase, see \cite{Berry1}. In recent papers, see
e.g.~\cite{Heiss2000, Dembowsky, Berry2, Dembowsky2003, KKM, KKM1,
Heiss, Stehmann}, two important cases are distinguished: the
diabolic points (DPs) and the exceptional points (EPs). From
mathematical point of view DP is a point where the eigenvalues
coalesce (become double), while corresponding eigenvectors remain
different (linearly independent); and EP is a point where both
eigenvalues and eigenvectors merge forming a Jordan block. Both
the DP and EP cases are interesting in applications and were
observed in experiments, see e.g.~\cite{Dembowsky, Dembowsky2003,
Stehmann}. In early studies only real and Hermitian matrices were
considered while modern physical systems require study of complex
symmetric and non-symmetric matrices, see \cite{MH, Berry2, KKM}.
Note that most of the cited papers dealt with specific $2\times2$
matrices depending on two or three parameters. Of course, in the
vicinity of an EP (and also DP) the $m$-dimensional matrix problem
becomes effectively two-dimensional, but finding the corresponding
two-dimensional space for a general $m$-dimensional matrix family
is a nontrivial problem \cite{Arnold2}.

In this paper we present a general theory of coupling of
eigenvalues of complex matrices of arbitrary dimension smoothly
depending on multiple real parameters. Two essential cases of weak
and strong coupling based on a Jordan form of the system matrix
are distinguished. These two cases correspond to diabolic and
exceptional points, respectively. We derive general formulae
describing coupling and decoupling of eigenvalues, crossing and
avoided crossing of eigenvalue surfaces. We present typical
(generic) pictures showing movement of eigenvalues, the eigenvalue
surfaces and their cross-sections. It is emphasized that the
presented theory of coupling of eigenvalues of complex matrices
gives not only qualitative, but also quantitative results on
behavior of eigenvalues based only on the information taken at the
singular points. Two examples on propagation of light in a
homogeneous non-magnetic crystal possessing natural optical
activity (chirality) and dichroism (absorption) in addition to
biaxial birefringence illustrate basic ideas and effectiveness of
the developed theory.

The presented theory is based on previous research on interaction
of eigenvalues of real matrices depending on multiple parameters
with mechanical applications. In \cite{Seyranian1991,
Seyranian1993} the important notion of weak and strong coupling
(interaction) was introduced for the first time. In the papers
\cite{SP1993, SLO1994, MS1999, SKliem2001, SM2001, KS2002, SM2003,
K2004, KS2004}, and in the recent book \cite{SeyMai2004}
significant mechanical effects related to diabolic and exceptional
points were studied. These include transference of instability
between eigenvalue branches, bimodal solutions in optimal
structures under stability constraints, flutter and divergence
instabilities in undamped nonconservative systems, effect of
gyroscopic stabilization, destabilization of a nonconservative
system by infinitely small damping, which were described and
explained from the point of view of coupling of eigenvalues. An
interesting application of the results on eigenvalue coupling to
electrical engineering problems is given in~\cite{Dobson}.

The paper is organized as follows. In Section 2 we present general
results on weak and strong coupling of eigenvalues of complex
matrices depending on parameters. These two cases correspond to
the study of eigenvalue behavior near diabolic and exceptional
points. Section 3 is devoted to crossing and avoided crossing of
eigenvalue surfaces near double eigenvalues with one and two
eigenvectors. Two physical examples are presented in Section 4,
and finally we end up with the conclusion in Section 5.

\section{Coupling of eigenvalues}

Let us consider the eigenvalue problem
    \begin{equation}
    \mathbf{A}\mathbf{u} = \lambda\mathbf{u}
    \label{eq1.1}
    \end{equation}
for a general $m\times m$ complex matrix $\mathbf{A}$ smoothly
depending on a vector of $n$ real parameters $\mathbf{p} =
(p_1,\ldots,p_n)$. Assume that, at $\mathbf{p} = \mathbf{p}_0$,
the eigenvalue coupling occurs, i.e., the matrix $\mathbf{A}_0 =
\mathbf{A}(\mathbf{p}_0)$ has an eigenvalue $\lambda_0$ of
multiplicity $2$ as a root of the characteristic equation
$\det(\mathbf{A}_0-\lambda_0\mathbf{I}) = 0$; $\mathbf{I}$ is the
identity matrix. This double eigenvalue can have one or two
linearly independent eigenvectors $\mathbf{u}$, which determine
the geometric multiplicity. The eigenvalue problem adjoint to
(\ref{eq1.1}) is
    \begin{equation}
    \mathbf{A}^*\mathbf{v} = \eta\mathbf{v},
    \label{eq1.1b}
    \end{equation}
where $\mathbf{A}^* = \overline{\mathbf{A}}^T$ is the adjoint
matrix operator (Hermitian transpose), see e.g. \cite{Lancaster}.
The eigenvalues $\lambda$ and $\eta$ of problems (\ref{eq1.1}) and
(\ref{eq1.1b}) are complex conjugate: $\eta = \overline{\lambda}$.

Double eigenvalues appear at sets in parameter space, whose
codimensions depend on the matrix type and the degeneracy (EP or
DP). In Table~\ref{tab1}, we list these codimensions based on the
results of the singularity theory~\cite{Neumann, Arnold2}. In this
paper we analyze general (nonsymmetric) complex matrices. The EP
degeneracy is the most typical for this type of matrices. In
comparison with EP, the DP degeneracy is a rare phenomenon in
systems described by general complex matrices. However, some
nongeneric situations may be interesting from the physical point
of view. As an example, we mention complex non-Hermitian
perturbations of symmetric two-parameter real matrices, when the
eigenvalue surfaces have coffee-filter singularity, see~\cite{MH,
Berry2, KKM}. A general theory of this phenomenon will be given in
our companion paper~\cite{KMS}.

\begin{table}
\centering
\begin{tabular}{|c|c|c|} \hline
  matrix type & codimension of DP & codimension of EP \\ \hline\hline
  real symmetric & 2 & non-existent \\ \hline
  real nonsymmetric & 3 & 1 \\ \hline
  Hermitian & 3 & non-existent \\ \hline
  complex symmetric & 4 & 2 \\ \hline
  complex nonsymmetric & 6 & 2 \\ \hline
\end{tabular}
\caption{Codimensions of eigenvalue degeneracies.}\label{tab1}
\end{table}

Let us consider a smooth perturbation of parameters in the form
$\mathbf{p} = \mathbf{p}(\varepsilon)$, where $\mathbf{p}(0) =
\mathbf{p}_0$ and $\varepsilon$ is a small real number. For the
perturbed matrix $\mathbf{A} =
\mathbf{A}(\mathbf{p}(\varepsilon))$, we have
    \begin{equation}
    \begin{array}{c}
        \mathbf{A} =
        \mathbf{A}_0+\varepsilon\mathbf{A}_1
        +\frac12\varepsilon^2\mathbf{A}_2+o(\varepsilon^2),\\[12pt]
        \displaystyle
        \mathbf{A}_0 = \mathbf{A}(\mathbf{p}_0),\quad
        \mathbf{A}_1
        = \sum_{i = 1}^n \frac{\partial\mathbf{A}}{\partial p_i}
        \frac{dp_i}{d\varepsilon},\quad
        \mathbf{A}_2
        = \sum_{i = 1}^n \frac{\partial\mathbf{A}}{\partial p_i}
        \frac{d^2p_i}{d\varepsilon^2}
        +\sum_{i,j = 1}^n
        \frac{\partial^2\mathbf{A}}{\partial p_i\partial p_j}
        \frac{dp_i}{d\varepsilon}\frac{dp_j}{d\varepsilon}.
    \end{array}
    \label{eq1.2}
    \end{equation}
The double eigenvalue $\lambda_0$ generally splits into a pair of
simple eigenvalues under the perturbation. Asymptotic formulae for
these eigenvalues and corresponding eigenvectors contain integer
or fractional powers of $\varepsilon$~\cite{Vishik}.

\subsection{Weak coupling of eigenvalues}

Let us consider the coupling of eigenvalues in the case of
$\lambda_0$ with two linearly independent eigenvectors
$\mathbf{u}_1$ and $\mathbf{u}_2$. This coupling point is known as
a diabolic point. Let us denote by $\mathbf{v}_1$ and
$\mathbf{v}_2$ two eigenvectors of the complex conjugate
eigenvalue $\eta = \overline{\lambda}$ for the adjoint eigenvalue
problem (\ref{eq1.1b}) satisfying the normalization conditions
    \begin{equation}
    (\mathbf{u}_1,\mathbf{v}_1) =
    (\mathbf{u}_2,\mathbf{v}_2) = 1,\quad
    (\mathbf{u}_1,\mathbf{v}_2) =
    (\mathbf{u}_2,\mathbf{v}_1) = 0,
    \label{eq1.NC}
    \end{equation}
where $(\mathbf{u},\mathbf{v}) = \sum_{i = 1}^n u_i\overline{v}_i$
denotes the Hermitian inner product. Conditions (\ref{eq1.NC})
define the unique vectors $\mathbf{v}_1$ and $\mathbf{v}_2$ for
given $\mathbf{u}_1$ and $\mathbf{u}_2$~\cite{SeyMai2004}.

For nonzero small $\varepsilon$, the two eigenvalues $\lambda_+$
and $\lambda_-$ resulting from the bifurcation of $\lambda_0$ and
the corresponding eigenvectors $\mathbf{u}_\pm$ are given by
    \begin{equation}
    \lambda_\pm
    = \lambda_0+\mu_\pm\varepsilon+o(\varepsilon),\quad
    \mathbf{u}_\pm
    = \alpha_\pm\mathbf{u}_1+\beta_\pm\mathbf{u}_2+o(1).
    \label{eq1.3}
    \end{equation}
The coefficients $\mu_\pm$, $\alpha_\pm$, and $\beta_\pm$ are
found from the $2\times 2$ eigenvalue problem \\ (see
e.g.~\cite{SeyMai2004})
    \begin{equation}
    \left(\begin{array}{cc}
        (\mathbf{A}_1\mathbf{u}_1,\mathbf{v}_1) &
        (\mathbf{A}_1\mathbf{u}_2,\mathbf{v}_1) \\[3pt]
        (\mathbf{A}_1\mathbf{u}_1,\mathbf{v}_2) &
        (\mathbf{A}_1\mathbf{u}_2,\mathbf{v}_2)
    \end{array}\right)
    \left(\begin{array}{cc}
        \alpha_\pm \\[2pt] \beta_\pm
    \end{array}\right)
    = \mu_\pm
    \left(\begin{array}{cc}
        \alpha_\pm \\[2pt] \beta_\pm
    \end{array}\right).
    \label{eq1.4}
    \end{equation}
Solving the characteristic equation for (\ref{eq1.4}), we find
    \begin{equation}
    \mu_\pm = \frac{(\mathbf{A}_1\mathbf{u}_1,\mathbf{v}_1)
    +(\mathbf{A}_1\mathbf{u}_2,\mathbf{v}_2)}{2}\pm
    \sqrt{\frac{((\mathbf{A}_1\mathbf{u}_1,\mathbf{v}_1)
    -(\mathbf{A}_1\mathbf{u}_2,\mathbf{v}_2))^2}{4}
    +(\mathbf{A}_1\mathbf{u}_1,\mathbf{v}_2)
    (\mathbf{A}_1\mathbf{u}_2,\mathbf{v}_1)}.
    \label{eq1.4b}
    \end{equation}
We note that for Hermitian matrices $\mathbf{A}$ one can take
$\mathbf{v}_1 = \mathbf{u}_1$ and $\mathbf{v}_2 = \mathbf{u}_2$ in
(\ref{eq1.4}), where the eigenvectors $\mathbf{u}_1$ and
$\mathbf{u}_2$ are chosen satisfying the conditions
$(\mathbf{u}_1,\mathbf{u}_1) = (\mathbf{u}_2,\mathbf{u}_2) = 1$
and $(\mathbf{u}_1,\mathbf{u}_2) = 0$, and obtain the well-known
formula, see~\cite{Hilbert}.

\begin{figure}
        \begin{center}
        \includegraphics[angle=0, width=0.95\textwidth]{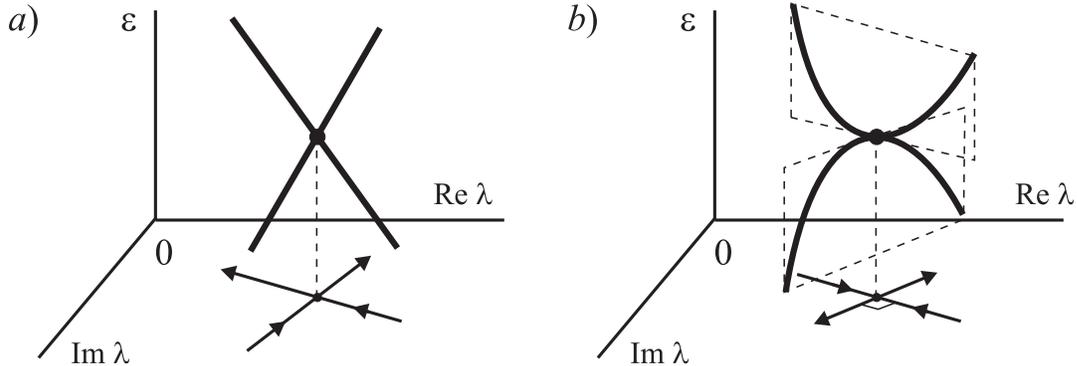}
        \end{center}
    \caption{Eigenvalue coupling: (a) weak, (b) strong.}
    \label{fig1}
    \end{figure}

As the parameter vector passes the coupling point $\mathbf{p}_0$
along the curve $\mathbf{p}(\varepsilon)$ in parameter space, the
eigenvalues $\lambda_+$ and $\lambda_-$ change smoothly and cross
each other at $\lambda_0$, see Figure~\ref{fig1}a. At the same
time, the corresponding eigenvectors $\mathbf{u}_+$ and
$\mathbf{u}_-$ remain different (linearly independent) at all
values of $\varepsilon$ including the point $\mathbf{p}_0$. We
call this interaction \textit{weak coupling}. By means of
eigenvectors, the eigenvalues $\lambda_\pm$ are well distinguished
during the weak coupling.

We emphasize that despite the eigenvalues $\lambda_\pm$ and the
eigenvectors $\mathbf{u}_\pm$ depend smoothly on a single
parameter $\varepsilon$, they are nondifferentiable functions of
multiple parameters at $\mathbf{p}_0$ in the sense of
Frech\'et~\cite{AnalysisTextbook}.

\subsection{Strong coupling of eigenvalues}

Let us consider coupling of eigenvalues at $\mathbf{p}_0$ with a
double eigenvalue $\lambda_0$ possessing a single eigenvector
$\mathbf{u}_0$. This case corresponds to the exceptional point.
The second vector of the invariant subspace corresponding to
$\lambda_0$ is called an associated vector $\mathbf{u}_1$ (also
called a generalized eigenvector~\cite{Lancaster}); it is
determined by the equation
    \begin{equation}
    \mathbf{A}_0\mathbf{u}_1 = \lambda_0\mathbf{u}_1+\mathbf{u}_0.
    \label{eq1.5}
    \end{equation}
An eigenvector $\mathbf{v}_0$ and an associated vector
$\mathbf{v}_1$ of the matrix $\mathbf{A}^*$ are determined by
    \begin{equation}
    \mathbf{A}_0^*\mathbf{v}_0 = \overline{\lambda}_0\mathbf{v}_0,\quad
    \mathbf{A}_0^*\mathbf{v}_1
    = \overline{\lambda}_0\mathbf{v}_1+\mathbf{v}_0,\quad
    (\mathbf{u}_1,\mathbf{v}_0) = 1,\quad
    (\mathbf{u}_1,\mathbf{v}_1) = 0,
    \label{eq1.6}
    \end{equation}
where the last two equations are the normalization conditions
determining $\mathbf{v}_0$ and $\mathbf{v}_1$ uniquely for a given
$\mathbf{u}_1$.

Bifurcation of $\lambda_0$ into two eigenvalues $\lambda_\pm$ and
the corresponding eigenvectors $\mathbf{u}_\pm$ are described by
(see e.g.~\cite{SeyMai2004})
    \begin{equation}
    \begin{array}{rcl}
        \lambda_\pm
        & = & \lambda_0\pm\sqrt{\mu_1\varepsilon}
        +\mu_2\varepsilon+o(\varepsilon),\\[5pt]
        \mathbf{u}_\pm
        & = & \mathbf{u}_0\pm\mathbf{u}_1\sqrt{\mu_1\varepsilon}
        +(\mu_1\mathbf{u}_0+\mu_2\mathbf{u}_1
        -\mathbf{G}^{-1}\mathbf{A}_1\mathbf{u}_0)\varepsilon
        +o(\varepsilon),
    \end{array}
    \label{eq1.7}
    \end{equation}
where $\mathbf{G} =
\mathbf{A}_0-\lambda_0\mathbf{I}+\mathbf{u}_1\mathbf{v}_1^*$. The
coefficients $\mu_1$ and $\mu_2$ are
    \begin{equation}
    \mu_1 = (\mathbf{A}_1\mathbf{u}_0,\mathbf{v}_0),\quad
    \mu_2 = \big((\mathbf{A}_1\mathbf{u}_0,\mathbf{v}_1)
    +(\mathbf{A}_1\mathbf{u}_1,\mathbf{v}_0)\big)/2.
    \label{eq1.8}
    \end{equation}

With a change of $\varepsilon$ from negative to positive values,
the two eigenvalues $\lambda_\pm$ approach, collide with infinite
speed (derivative with respect to $\varepsilon$ tends to infinity)
at $\lambda_0$, and diverge in the perpendicular direction, see
Figure~\ref{fig1}b. The eigenvectors interact too. At $\varepsilon
= 0$, they merge to $\mathbf{u}_0$ up to a scalar complex factor.
At nonzero $\varepsilon$, the eigenvectors $\mathbf{u}_\pm$ differ
from $\mathbf{u}_0$ by the leading term
$\pm\mathbf{u}_1\sqrt{\mu_1\varepsilon}$. This term takes the
purely imaginary factor $i$ as $\varepsilon$ changes the sign, for
example altering from negative to positive values.

We call such a coupling of eigenvalues as \textit{strong}. An
exciting feature of the strong coupling is that the two
eigenvalues cannot be distinguished after the interaction. Indeed,
there is no natural rule telling how the eigenvalues before
coupling correspond to those after the coupling.

\section{Crossing of eigenvalue surfaces}

\subsection{Double eigenvalue with single eigenvector}

Let, at the point ${\bf p}_0$, the spectrum of the complex matrix
family ${\bf A}({\bf p})$ contain a double complex eigenvalue
$\lambda_0$ with an eigenvector ${\bf u}_0$ and an associated
vector ${\bf u}_1$. The splitting of the double eigenvalue with a
change of the parameters is governed by equations (\ref{eq1.7})
and (\ref{eq1.8}). Introducing the real $n$-dimensional vectors
$\bf f$, $\bf g$, $\bf h$, $\bf r$ with the components
    \begin{equation}
    {f}_s={\rm Re}\left(\frac{\partial \bf A}{\partial p_s} {\bf u}_0, {\bf v}_0\right),~~
    {g}_s={\rm Im}\left(\frac{\partial \bf A}{\partial p_s} {\bf u}_0, {\bf v}_0\right),
    \label{eq2.1}
    \end{equation}
    \begin{equation}
    {h}_s = {\rm Re}
    \left(\left(\frac{\partial \bf A}{\partial p_s} {\bf u}_0, {\bf
    v}_1\right)+
    \left(\frac{\partial \bf A}{\partial p_s} {\bf u}_1, {\bf v}_0\right)\right),~~
    {r}_s = {\rm Im}
    \left(\left(\frac{\partial \bf A}{\partial p_s} {\bf u}_0, {\bf v}_1\right)+
    \left(\frac{\partial \bf A}{\partial p_s} {\bf u}_1, {\bf v}_0\right)\right),
    \label{eq2.1a}
    \end{equation}
    $$
    s=1,\ldots,n.
    $$
and neglecting higher order terms, we obtain from (\ref{eq1.7}) an
asymptotic formula
    \begin{equation}
    {\rm Re}\Delta\lambda+i{\rm Im}\Delta\lambda=
    \pm\sqrt{\langle {\bf f}, \Delta {\bf p} \rangle+
    i\langle {\bf g},\Delta {\bf p}\rangle}+
    \frac{1}{2}(\langle {\bf h},\Delta {\bf p}\rangle+
    i\langle {\bf r},\Delta {\bf p} \rangle),
    \label{eq2.2}
    \end{equation}
where $\Delta \lambda = \lambda_{\pm}-\lambda_0$, $\Delta{\bf p} =
{\bf p}-{\bf p}_0$, and angular brackets denote inner product of
real vectors: $\langle {\bf a}, {\bf b}
\rangle=\sum_{s=1}^na_sb_s$. From equation (\ref{eq2.2}) it is
clear that the eigenvalue remains double in the first
approximation if the two following equations are satisfied
    \begin{equation}\label{eq2.2a}
    \langle {\bf f}, \Delta {\bf p}\rangle=0,~~
    \langle {\bf g},\Delta {\bf p} \rangle=0.
    \end{equation}
This means that the double complex eigenvalue with the Jordan
chain of length 2 has codimension 2. Thus, double complex
eigenvalues occur at isolated points of the plane of two
parameters, and in the three-parameter space the double
eigenvalues form a curve \cite{Arnold2}. Equations (\ref{eq2.2a})
define a tangent line to this curve at the point ${\bf p}_0$.

Taking square of (\ref{eq2.2}), where the terms linear with
respect to the increment of parameters are neglected, and
separating real and imaginary parts, we derive the equations
    \begin{equation}
    ({\rm Re}\Delta \lambda)^2-({\rm Im}\Delta \lambda)^2=
    \langle {\bf f},\Delta {\bf p} \rangle,~~
    2{\rm Re}\Delta \lambda\,{\rm Im}\Delta\lambda=
    \langle {\bf g},\Delta {\bf p} \rangle.
    \label{eq2.3}
    \end{equation}
Let us assume that $f_1^2+g_1^2 \ne 0$, which is the nondegeneracy
condition for the complex eigenvalue $\lambda_0$. Isolating the
increment $\Delta p_1$ in one of the equations (\ref{eq2.3}) and
substituting it into the other one we get
    \begin{equation}
    g_1({\rm Re}\Delta\lambda)^2-
    2f_1{\rm Re}\Delta\lambda\,{\rm Im}\Delta\lambda-
    g_1({\rm Im}\Delta\lambda)^2=\gamma,
    \label{eq2.4}
    \end{equation}
where $\gamma$ is a small real constant
    \begin{equation}
    \gamma=\sum_{s=2}^n(f_sg_1-f_1g_s)\Delta p_s.
    \label{eq2.5}
    \end{equation}
Equation (\ref{eq2.4}) describes hyperbolic trajectories of the
eigenvalues $\lambda_{\pm}$ in the complex plane when only $\Delta
p_1$ is changed and the increments $\Delta p_2$, $\ldots$, $\Delta
p_n$ are fixed. Of course, any component of the vector $\Delta
{\bf p}$ can be chosen instead of $\Delta p_1$.

\begin{figure}
        \begin{center}
        \includegraphics[angle=0, width=0.8\textwidth]{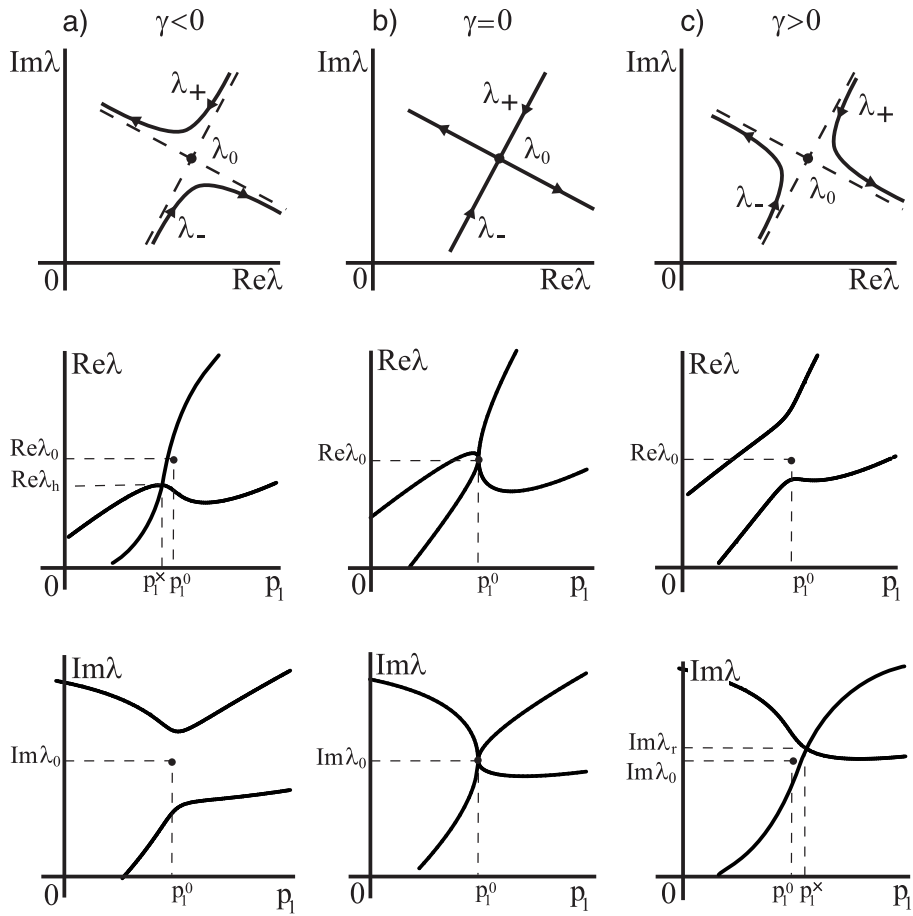}
        \end{center}
    \caption{Crossing and avoided crossing of eigenvalues.}
    \label{fig2}
    \end{figure}

Let us study movement of eigenvalues in the complex plane in more
detail. If $\Delta p_j=0$, $j=2,\ldots,n$, or if they are nonzero
but satisfy the equality $\gamma=0$, then equation (\ref{eq2.4})
yields two perpendicular lines which for $g_1 \ne 0$ are described
by the expression
    \begin{equation}
    g_1{\rm Re}(\lambda-\lambda_0)-
    \left(f_1\pm\sqrt{f_1^2+g_1^2}\right)
    {\rm Im}(\lambda-\lambda_0)=0.
    \label{eq2.6}
    \end{equation}
These lines intersect at the point $\lambda_0$ of the complex
plane. Due to variation of the parameter $p_1$ two eigenvalues
$\lambda_{\pm}$ approach along one of the lines (\ref{eq2.6}),
merge to $\lambda_0$ at $\Delta p_1 = 0 $, and then diverge along
the other line (\ref{eq2.6}), perpendicular to the line of
approach; see Figure~\ref{fig2}b, where the arrows show motion of
eigenvalues with a monotonous change of $p_1$. Recall that the
eigenvalues that born after the coupling cannot be identified with
the eigenvalues before the coupling.

If $\gamma \ne 0$, then equation (\ref{eq2.4}) defines a hyperbola
in the complex plane. Indeed, for $g_1 \ne 0$ it is transformed to
the equation of hyperbola
    \begin{equation}
    (g_1{\rm Re}(\lambda-\lambda_0)-
    f_1{\rm Im}(\lambda-\lambda_0))^2-
    ({\rm Im}(\lambda-\lambda_0))^2(f_1^2+g_1^2)=
    \gamma g_1
    \label{eq2.7}
    \end{equation}
with the asymptotes described by equation (\ref{eq2.6}). As
$\Delta p_1$ changes monotonously, two eigenvalues $\lambda_+$ and
$\lambda_-$ moving each along its own branch of hyperbola come
closer, turn and diverge; see Figure~\ref{fig2}a,c. Note that for
a small $\gamma$ the eigenvalues $\lambda_{\pm}$ come arbitrarily
close to each other without coupling that means {\it avoided
crossing}. When $\gamma$ changes the sign, the quadrants
containing hyperbola branches are changed to the adjacent.

Expressing ${\rm Im}\Delta\lambda$ from the second of equations
(\ref{eq2.3}), substituting it into the first equation and then
isolating ${\rm Re}\Delta\lambda$, we find
    \begin{equation}
    {\rm Re}\lambda_{\pm}=\lambda_0+\frac{1}{2}\langle {\bf h}, \Delta {\bf p}\rangle\pm
    \sqrt{\frac{1}{2}\left(\langle {\bf f}, \Delta {\bf p}\rangle+
    \sqrt{\langle {\bf f},\Delta {\bf p}\rangle^2+\langle {\bf g},\Delta {\bf p}\rangle^2}\right)}.
    \label{eq2.8}
    \end{equation}
Similar transformation yields
    \begin{equation}
    {\rm Im}\lambda_{\pm}=\lambda_0+\frac{1}{2}\langle {\bf r}, \Delta {\bf p}\rangle\pm
    \sqrt{\frac{1}{2}\left(-\langle {\bf f}, \Delta {\bf
    p}\rangle+
    \sqrt{\langle {\bf f},\Delta {\bf p}\rangle^2+\langle {\bf g},\Delta {\bf p}\rangle^2}\right)}.
    \label{eq2.9}
    \end{equation}

Equations (\ref{eq2.8}) and (\ref{eq2.9}) describe behavior of
real and imaginary parts of eigenvalues $\lambda_{\pm}$ with a
change of the parameters. On the other hand they define
hypersurfaces in the spaces $(p_1,p_2,\ldots,p_n,{\rm Re}\lambda)$
and $(p_1,p_2,\ldots,p_n,{\rm Im}\lambda)$. The sheets ${\rm
Re}\lambda_+(\bf p)$ and ${\rm Re}\lambda_-(\bf p)$ of the
eigenvalue hypersurface (\ref{eq2.8}) are connected at the points
of the set
    \begin{equation}\label{eq2.10}
    {\rm Re}\Delta\lambda=\frac{1}{2}\langle {\bf h}, \Delta {\bf p}\rangle,~~
    \langle {\bf g}, \Delta {\bf p}\rangle=0,~~\langle {\bf f}, \Delta {\bf
    p}\rangle \le 0,
    \end{equation}
where the real parts of the eigenvalues $\lambda_{\pm}$ coincide:
${\rm Re}\lambda_-={\rm Re}\lambda_+$. Similarly, the set
    \begin{equation}\label{eq2.11}
    {\rm Im}\Delta\lambda=\frac{1}{2}\langle {\bf r}, \Delta {\bf p}\rangle,~~
    \langle {\bf g}, \Delta {\bf p}\rangle=0,~~\langle {\bf f}, \Delta {\bf
    p}\rangle \ge 0,
   \end{equation}
glues the sheets ${\rm Im}\lambda_+(\bf p)$ and ${\rm
Im}\lambda_-(\bf p)$ of the eigenvalue hypersurface (\ref{eq2.9}).

To study the geometry of the eigenvalue hypersurfaces we look at
their two-dimensional cross-sections. Consider for example the
functions ${\rm Re}\lambda(p_1)$ and ${\rm Im}\lambda(p_1)$ at
fixed values of the other parameters $p_2,p_3,\ldots,p_n$. When
the increments $\Delta p_s=0$, $s=2,3,\ldots,n$, both the real and
imaginary parts of the eigenvalues $\lambda_{\pm}$ cross at
$p_1=p_1^0$, see Figure~\ref{fig2}b. The crossings are described
by the double cusps defined by the equations following from
(\ref{eq2.8}) and (\ref{eq2.9}) as
    \begin{equation}\label{eq2.12}
    {\rm Re}\Delta\lambda =
    \pm\sqrt{
    \frac{f_1\pm\sqrt{f_1^2+g_1^2}}{2}\Delta p_1}+\frac{h_1}{2}\Delta p_1,
    ~~
    {\rm Im}\Delta\lambda =
    \pm\sqrt{
    \frac{-f_1\pm\sqrt{f_1^2+g_1^2}}{2}\Delta p_1}+\frac{r_1}{2}\Delta p_1.
    \end{equation}

For the fixed $\Delta p_s\ne0$, $s=2,3,\ldots,n$, either real
parts of the eigenvalues $\lambda_{\pm}$ cross due to variation of
$p_1$ while the imaginary parts avoid crossing or vice-versa, as
shown in Figure~\ref{fig2}a,c. Note that these two cases
correspond to level crossing and width repulsion or vice-versa
studied in \cite{Heiss2000}. The crossings, which occur at
$p_1^{\times}=p_1^0-\sum_{s=2}^n (g_s/g_1)\Delta p_s$ and
\begin{equation}\label{eq2.15}
{\rm Re}\lambda_h={\rm
Re}\lambda_0-\frac{1}{2g_1}\sum_{s=2}^n(h_1g_s-g_1h_s)\Delta
p_s,~~{\rm Im}\lambda_r={\rm
Im}\lambda_0-\frac{1}{2g_1}\sum_{s=2}^n(r_1g_s-g_1r_s)\Delta p_s,
\end{equation}
are described by the equations (\ref{eq2.8}) and (\ref{eq2.9}). In
the vicinity of the crossing points the tangents of two
intersecting curves are
    \begin{equation}\label{eq2.13}
    {\rm Re}\lambda={\rm Re}\lambda_h+
    \left(\frac{h_1}{2}\pm\frac{g_1}{2}\sqrt{\frac{g_1}{\gamma}} \right)(p_1-p_1^{\times}),
    \end{equation}
    \begin{equation}\label{eq2.14}
    {\rm Im}\lambda={\rm Im}\lambda_r+
    \left(\frac{r_1}{2}\pm\frac{g_1}{2}\sqrt{-\frac{g_1}{\gamma}} \right)(p_1-p_1^{\times}),
    \end{equation}
where the coefficient $\gamma$ is defined by equation
(\ref{eq2.5}). Lines (\ref{eq2.13}) and (\ref{eq2.14}) tend to the
vertical position as $\gamma\rightarrow 0$ and coincide at
$\gamma=0$. The avoided crossings are governed by the equations
(\ref{eq2.8}) and (\ref{eq2.9}).

    \begin{figure}
        \begin{center}
        \includegraphics[angle=0, width=0.85\textwidth]{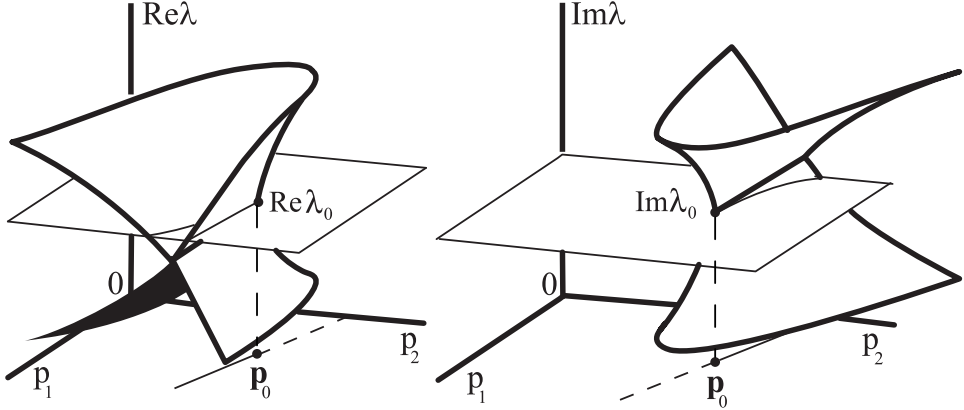}
        \end{center}
    \caption{Crossing of eigenvalue surfaces near
    the double eigenvalue with single eigenvector.}
    \label{fig3}
    \end{figure}

If the vector of parameters consists of only two components ${\bf
p}=(p_1, p_2)$, then in the vicinity of the point ${\bf p}_0$,
corresponding to the double eigenvalue $\lambda_0$, the eigenvalue
surfaces (\ref{eq2.8}) and (\ref{eq2.9}) have the form of the
well-known Whitney umbrella; see Figure~\ref{fig3}. The sheets of
the eigensurfaces are connected along the rays (\ref{eq2.10}) and
(\ref{eq2.11}). We emphasize that these rays are inclined with
respect to the plane of the parameters $p_1$, $p_2$. The
cross-sections of the eigensurfaces by the planes orthogonal to
the axis $p_2$, described by the equations
(\ref{eq2.12})--(\ref{eq2.14}), are shown in Figure~\ref{fig2}.

Note that the rays (\ref{eq2.10}), (\ref{eq2.11}) and the point
${\bf p}_0$ are well-known in crystal optics as {\it the branch
cuts} and {\it the singular axis}, respectively \cite{Berry2}. We
emphasize that the branch cut is a general phenomenon: it always
appears near the EP degeneracy. In general, branch cuts may be
infinite or end up at another EP. The second scenario is always
the case when the complex matrix $\mathbf{A}(\mathbf{p})$ is a
small perturbation of a family of real symmetric
matrices~\cite{KMS}.

Consider the movement of the eigenvalues in the complex plane near
the point ${\bf p}_0$ due to cyclic variation of the parameters
$p_1$ and $p_2$ of the form $\Delta p_1=a+r\cos{\varphi}$ and
$\Delta p_2=b+r\sin{\varphi}$, where $a$, $b$, and $r$ are small
parameters of the same order. From equations (\ref{eq2.3}) we
derive
    $$
    (g_1{\rm Re}\Delta\lambda^2-2f_1{\rm Re}\Delta\lambda{\rm Im}\Delta\lambda-
     g_1{\rm Im}\Delta\lambda^2-b(f_2g_1-f_1g_2))^2+
    $$
    \begin{equation}\label{eq2.16}
   +(g_2{\rm Re}\Delta\lambda^2-2f_2{\rm Re}\Delta\lambda{\rm Im}\Delta\lambda-
     g_2{\rm Im}\Delta\lambda^2-a(f_1g_2-g_1f_2))^2=
    (f_2g_1-f_1g_2)^2r^2.
    \end{equation}

    \begin{figure}
    \begin{center}
    \includegraphics[angle=0, width=1\textwidth]{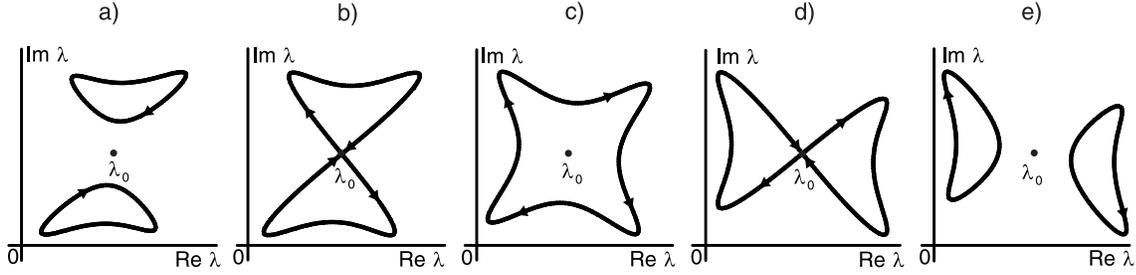}
    \end{center}
    \caption{Movement of eigenvalues due to cyclic evolution of the parameters.}
    \label{fig4}
    \end{figure}

Movement of eigenvalues on the complex plane governed by equation
(\ref{eq2.16}) is shown in Figure~\ref{fig4}. If the contour
encircles the point ${\bf p}_0$, then the eigenvalues move along
the curve (\ref{eq2.16}) around the double eigenvalue $\lambda_0$
in the complex plane, see Figure~\ref{fig4}c. Indeed, in this case
$a^2+b^2<r^2$ and the loop (\ref{eq2.16}) crosses the lines ${\rm
Re}\lambda={\rm Re}\lambda_0$ and ${\rm Im}\lambda={\rm
Im}\lambda_0$ at the four points given by the equations
    \begin{equation}\label{eq2.18}
    ({\rm Im}\Delta\lambda)^2 = \frac{(f_2g_1-f_1g_2)\left(g_2 a -g_1 b\pm \sqrt{(g_2
    a-g_1
    b)^2+(r^2-a^2-b^2)(g_1^2+g_2^2)}\right)}{g_1^2+g_2^2}
    \end{equation}
and
    \begin{equation}\label{eq2.17}
    ({\rm Re}\Delta\lambda)^2 = \frac{(f_2g_1-f_1g_2)
    \left(g_1 b -g_2 a \pm \sqrt{(g_1 b - g_2
    a)^2+(r^2-a^2-b^2)(g_1^2+g_2^2)}\right)}{g_1^2+g_2^2},
    \end{equation}
respectively. When $a^2+b^2=r^2$ the loop overlaps at the double
eigenvalue and its form depends on the sign of the quantity
$\sigma=(f_2g_1-f_1g_2)(g_1b-g_2a)$. If $\sigma<0$ the eigenvalues
cross the line ${\rm Re}\lambda={\rm Re}\lambda_0$
(Figure~\ref{fig4}b), otherwise they cross the line ${\rm
Im}\lambda={\rm Im}\lambda_0$ (Figure~\ref{fig4}d). Eigenvalues
strongly couple at the point $\lambda_0$ in the complex plane. For
$a^2+b^2>r^2$ the circuit in the parameter plane does not contain
the point ${\bf p}_0$ and the eigenvalues move along the two
different closed paths ("kidneys", \cite{Arnold1989}) in the
complex plane, see Figure~\ref{fig4}a,e. Each eigenvalue crosses
the line ${\rm Re}\lambda={\rm Re}\lambda_0$ twice for $\sigma<0$
(Figure~\ref{fig4}a), and for $\sigma>0$ they cross the line ${\rm
Im}\lambda={\rm Im}\lambda_0$ (Figure~\ref{fig4}d). Note that the
"kidneys" in the complex plane were observed in \cite{KKM1} for
the specific problem of Stark resonances for a double $\delta$
quantum well.

\subsection{Double eigenvalue with two eigenvectors}

Let $\lambda_0$ be a double eigenvalue of the matrix $\mathbf{A}_0
= \mathbf{A}(\mathbf{p}_0)$ with two eigenvectors $\mathbf{u}_1$
and $\mathbf{u}_2$. Under perturbation of parameters $\mathbf{p} =
\mathbf{p}_0+\Delta\mathbf{p}$, the bifurcation of $\lambda_0$
into two simple eigenvalues $\lambda_+$ and $\lambda_-$ occurs.
Using (\ref{eq1.3}) and (\ref{eq1.4b}), we obtain the asymptotic
formula for $\lambda_\pm$ under multiparameter perturbation as
    \begin{equation}
    \lambda_\pm
    = \lambda_0+\frac{\langle\mathbf{d}_{11}+\mathbf{d}_{22},
    \Delta\mathbf{p}\rangle}{2}\pm
    \sqrt{\frac{\langle\mathbf{d}_{11}-\mathbf{d}_{22},
    \Delta\mathbf{p}\rangle^2}{4}+\langle\mathbf{d}_{12},
    \Delta\mathbf{p}\rangle\langle\mathbf{d}_{21},
    \Delta\mathbf{p}\rangle},
    \label{eq3.1}
    \end{equation}
where $\mathbf{d}_{ij} = (d_{ij}^1,\ldots,d_{ij}^n)$ is a complex
vector with the components
    \begin{equation}
    d_{ij}^k = \left(\frac{\partial\mathbf{A}}{\partial p_k}
    \mathbf{u}_i,\mathbf{v}_j\right),
    \label{eq3.2}
    \end{equation}
and $\langle\mathbf{d}_{ij},\Delta\mathbf{p}\rangle =
\langle\mathrm{Re}\,\mathbf{d}_{ij},\Delta\mathbf{p}\rangle
+i\langle\mathrm{Im}\,\mathbf{d}_{ij},\Delta\mathbf{p}\rangle$. In
the same way as we derived formulae (\ref{eq2.8}) and
(\ref{eq2.9}), we obtain from (\ref{eq3.1}) the expressions for
real and imaginary parts of $\lambda_\pm$ in the form
    \begin{equation}
    \mathrm{Re}\,\lambda_\pm
    = \mathrm{Re}\,\lambda_0
    +\mathrm{Re}\,\langle\mathbf{d}_{11}+\mathbf{d}_{22},
    \Delta\mathbf{p}\rangle/2\pm
    \sqrt{(|c|+\mathrm{Re}\,c)/2},
    \label{eq3.3}
    \end{equation}
    \begin{equation}
    \mathrm{Im}\,\lambda_\pm
    = \mathrm{Im}\,\lambda_0
    +\mathrm{Im}\,\langle\mathbf{d}_{11}+\mathbf{d}_{22},
    \Delta\mathbf{p}\rangle/2\pm
    \sqrt{(|c|-\mathrm{Re}\,c)/2},
    \label{eq3.4}
    \end{equation}
where
    \begin{equation}
    c = \langle\mathbf{d}_{11}-\mathbf{d}_{22},
    \Delta\mathbf{p}\rangle^2/4+\langle\mathbf{d}_{12},
    \Delta\mathbf{p}\rangle\langle\mathbf{d}_{21},
    \Delta\mathbf{p}\rangle.
    \label{eq3.5}
    \end{equation}

Considering the situation when $\lambda_0$ remains double under
perturbation of parameters, i.e. $\lambda_+ = \lambda_-$, we
obtain the two independent equations
    \begin{equation}
    \mathrm{Re}\,c = 0,\quad\mathrm{Im}\,c = 0.
    \label{eq3.6}
    \end{equation}
By using (\ref{eq1.3})--(\ref{eq1.4b}), one can show that the
perturbed double eigenvalue $\lambda_+ = \lambda_-$ possesses a
single eigenvector $\mathbf{u}_+ = \mathbf{u}_-$, i.e., the weak
coupling becomes strong due to perturbation, see
\cite{SeyMai2004}.

The perturbed double eigenvalue has two eigenvectors only when the
matrix in the left-hand side of (\ref{eq1.4}) is proportional to
the identity matrix. This yields the equations
    \begin{equation}
    \langle\mathbf{d}_{11},\Delta\mathbf{p}\rangle
    = \langle\mathbf{d}_{22},\Delta\mathbf{p}\rangle,\quad
    \langle\mathbf{d}_{12},\Delta\mathbf{p}\rangle
    = \langle\mathbf{d}_{21},\Delta\mathbf{p}\rangle = 0,
    \label{eq3.7}
    \end{equation}
Conditions (\ref{eq3.7}) imply (\ref{eq3.6}) and represent six
independent equations taken for real and imaginary parts. Thus,
weak coupling of eigenvalues is a phenomenon of codimension 6,
which generically occurs at isolated points in 6-parameter space,
see~\cite{Arnold2, MH}.

\begin{figure}
        \begin{center}
        \includegraphics[angle=0, width=0.8\textwidth]{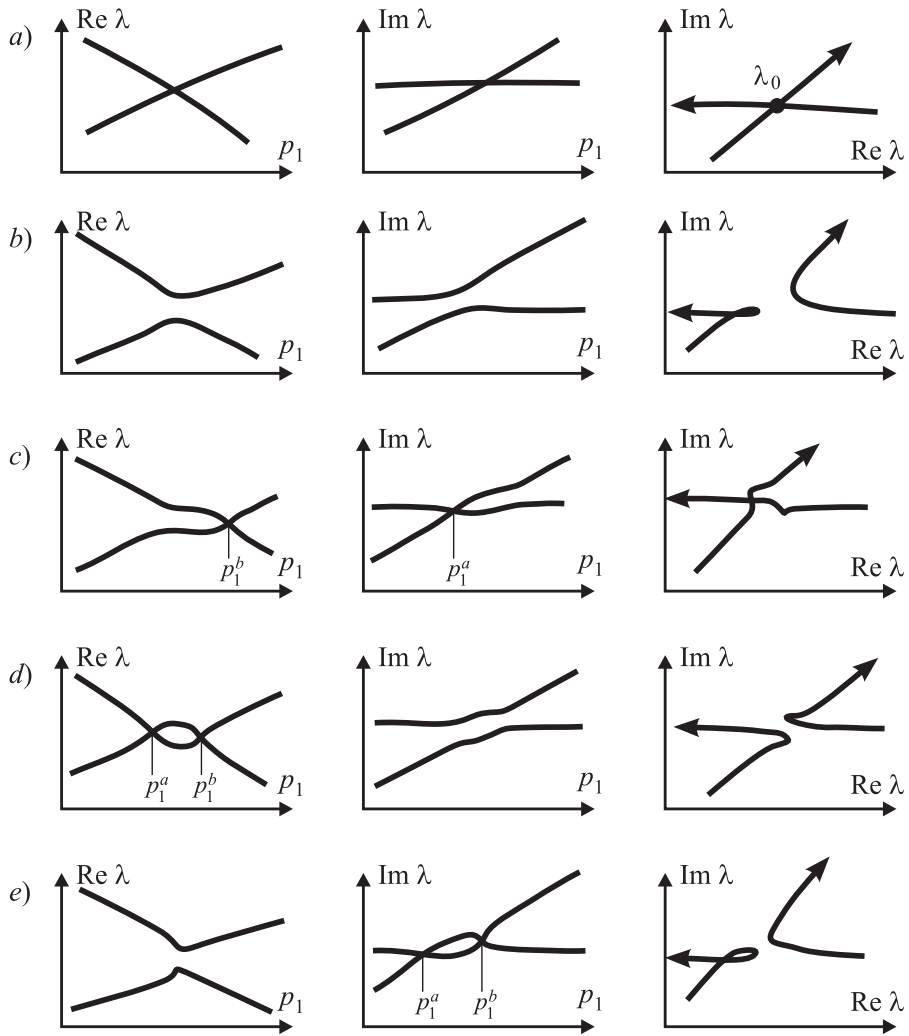}
        \end{center}
    \caption{Weak coupling of eigenvalues and avoided
crossing.}
    \label{fig3.1}
    \end{figure}

First, let us study behavior of the eigenvalues $\lambda_+$ and
$\lambda_-$ depending on one parameter, say $p_1$, when the other
parameters $p_2,\ldots,p_n$ are fixed in the neighborhood of the
coupling point $\lambda_+(\mathbf{p}_0) = \lambda_-(\mathbf{p}_0)
= \lambda_0$. In case $\Delta p_2 = \cdots = \Delta p_n = 0$,
expression (\ref{eq3.1}) yields
    \begin{equation}
    \lambda_\pm
    = \lambda_0+\left(\frac{d_{11}^1+d_{22}^1}{2}
    \pm\sqrt{\frac{(d_{11}^1-d_{22}^1)^2}{4}+d_{12}^1d_{21}^1}
    \right)\Delta p_1.
    \label{eq3.8}
    \end{equation}
The two eigenvalues couple when $\Delta p_1 = 0$ with the double
eigenvalue $\lambda_0$, see Figure~\ref{fig3.1}a. As we showed in
Section~2, the eigenvalues $\lambda_+$ and $\lambda_-$ behave as
smooth functions at the coupling point; they possess different
eigenvectors, which are smooth functions of $\Delta p_1$ too.

If the perturbations $\Delta p_2,\ldots,\Delta p_n$ are nonzero,
the avoided crossing of the eigenvalues $\lambda_\pm$ with a
change of $p_1$ is a typical scenario. We can distinguish
different cases by checking intersections of real and imaginary
parts of $\lambda_+$ and $\lambda_-$. By using (\ref{eq3.3}), we
find that $\mathrm{Re}\,\lambda_+ = \mathrm{Re}\,\lambda_-$ if
    \begin{equation}
    \mathrm{Im}\,c = 0,\quad \mathrm{Re}\,c < 0.
    \label{eq3.9}
    \end{equation}
Analogously, from (\ref{eq3.4}) it follows that
$\mathrm{Im}\,\lambda_+ = \mathrm{Im}\,\lambda_-$ if
    \begin{equation}
    \mathrm{Im}\,c = 0,\quad \mathrm{Re}\,c > 0.
    \label{eq3.10}
    \end{equation}

Let us write expression (\ref{eq3.5}) in the form
    \begin{equation}
    c = c_0+c_1\Delta p_1+c_2(\Delta p_1)^2,
    \label{eq3.11}
    \end{equation}
where
    \begin{equation}
    \begin{array}{c}
        c_0 = \displaystyle\sum_{k,l = 2}^n
        \left[(d_{11}^k-d_{22}^k)(d_{11}^l-d_{22}^l)/4
        +d_{12}^kd_{21}^l\right]\Delta p_k\Delta p_l, \\[15pt]
        c_1 = \displaystyle\sum_{k = 2}^n
        \left[(d_{11}^1-d_{22}^1)(d_{11}^k-d_{22}^k)/2
        +(d_{12}^1d_{21}^k+d_{12}^kd_{21}^1)\right]\Delta p_k,
        \\[12pt]
        c_2 =
        (d_{11}^1-d_{22}^1)^2/4+d_{12}^1d_{21}^1.
    \end{array}
    \label{eq3.12}
    \end{equation}
If the discriminant $D =
(\mathrm{Im}\,c_1)^2-4\mathrm{Im}\,c_0\mathrm{Im}\,c_2
> 0$, the equation $\mathrm{Im}\,c = 0$ yields two solutions
    \begin{equation}
    \Delta p_1^a = \frac{-\mathrm{Im}\,c_1-\sqrt{D}}{2\mathrm{Im}\,c_2},\quad
    \Delta p_1^b = \frac{-\mathrm{Im}\,c_1+\sqrt{D}}{2\mathrm{Im}\,c_2}.
    \label{eq3.13}
    \end{equation}
There are no real solutions if $D < 0$, and the single solution
corresponds to the degenerate case $D = 0$. At the points $p_1^a =
p_1^0+\Delta p_1^a$ and $p_1^b = p_1^0+\Delta p_1^b$ the values of
$c$ are real, and we denote them by $c_a$ and $c_b$, respectively.
According to (\ref{eq3.9}) and (\ref{eq3.10}), the sign of
$c_{a,b}$ determines whether the real or imaginary parts of
$\lambda_\pm$ coincide at $p_1^{a,b}$.

In the nondegenerate case $D \ne 0$, there are four types of
avoided crossing shown in Figure~\ref{fig3.1}b--e. The first case
corresponds to $D < 0$ when both real and imaginary parts of the
eigenvalues $\lambda_\pm$ are separate at all $p_1$, see
Figure~\ref{fig3.1}b. In other cases $D > 0$, so that there are
two separate points $p_{1}^a$ and $p_{1}^b$. For the second type
we have $c_a > 0$ and $c_b < 0$, when both real and imaginary
parts of $\lambda_\pm$ have a single intersection, see
Figure~\ref{fig3.1}c. The equivalent situation when $c_a < 0$ and
$c_b > 0$ is obtained by interchanging the points $p_{1}^a$ and
$p_{1}^b$ in Figure~\ref{fig3.1}c. The third type is represented
by $c_{a,b} < 0$, when the real parts of $\lambda_\pm$ have two
intersections and $\mathrm{Im}\,\lambda_\pm$ do not intersect, see
Figure~\ref{fig3.1}d. Finally, if $c_{a,b} > 0$, when the real
parts of $\lambda_\pm$ do not intersect and
$\mathrm{Im}\,\lambda_\pm$ intersect at both $p_{1}^a$ and
$p_{1}^b$, see Figure~\ref{fig3.1}e. The last column in
Figure~\ref{fig3.1} shows behavior of the eigenvalues
$\lambda_\pm$ on the complex plane. In each of the cases b--e, the
trajectories of eigenvalues on the complex plane may intersect
and/or self-intersect, which can be studied by using expression
(\ref{eq3.1}). Note that intersections of the eigenvalue
trajectories on the complex plane do not imply eigenvalue coupling
since the eigenvalues $\lambda_+$ and $\lambda_-$ pass the
intersection point at different values of $p_1$. The small loops
of the eigenvalue trajectories on the complex plane, shown in
Figure~\ref{fig3.1}b,e, shrink as the perturbations of the
parameters $\Delta p_2,\Delta p_3,\ldots,\Delta p_n$ tend to zero.
Finally, we mention that the case of Figure~\ref{fig3.1}c is the
only avoided crossing scenario when the eigenvalues follow the
initial directions on the complex plane after interaction. In the
other three cases (b,d, and e) the eigenvalues interchange their
directions due the interaction.

Let us consider a system depending on two parameters $p_1$ and
$p_2$ with the weak coupling of eigenvalues at $p_1 = p_1^0$ and
$p_2 = p_2^0$. The double eigenvalue $\lambda_0$ bifurcates into a
pair $\lambda_\pm$ under perturbation of the parameters $\Delta
p_1$ and $\Delta p_2$. Conditions (\ref{eq3.9}) and (\ref{eq3.10})
determine the values of parameters, at which the real and
imaginary parts of $\lambda_\pm$ coincide.

Let us write expression (\ref{eq3.5}) in the form
    \begin{equation}
    c = c_{11}(\Delta p_1)^2+c_{12}\Delta p_1\Delta p_2+c_{22}(\Delta p_2)^2,
    \label{eq3.14}
    \end{equation}
where
    \begin{equation}
    \begin{array}{c}
        c_{11} =
        (d_{11}^1-d_{22}^1)^2/4+d_{12}^1d_{21}^1,\quad
        c_{22} =
        (d_{11}^2-d_{22}^2)^2/4+d_{12}^2d_{21}^2,\\[5pt]
        c_{12} =
        (d_{11}^1-d_{22}^1)(d_{11}^2-d_{22}^2)/2+d_{12}^1d_{21}^2
        +d_{12}^2d_{21}^1.
    \end{array}
    \label{eq3.15}
    \end{equation}
If the discriminant $D' =
(\mathrm{Im}\,c_{12})^2-4\mathrm{Im}\,c_{11}\mathrm{Im}\,c_{22}
> 0$, the equation $\mathrm{Im}\,c = 0$ yields the two crossing lines
    \begin{equation}
    \begin{array}{l}
        l_a: \ \
        2\mathrm{Im}\,c_{11}\Delta p_1+(\mathrm{Im}\,c_{12}+\sqrt{D'})\Delta p_2 =
        0,\\[5pt]
        l_b: \ \
        2\mathrm{Im}\,c_{11}\Delta p_1+(\mathrm{Im}\,c_{12}-\sqrt{D'})\Delta p_2 =
        0.
    \end{array}
    \label{eq3.16}
    \end{equation}
There are no real solutions if $D' < 0$, and the lines $l_a$ and
$l_b$ coincide in the degenerate case $D' = 0$. On the lines
$l_{a,b}$ the values of $c$ are real numbers of the same sign; we
denote $\gamma_a = \mathrm{sign}\,c$ for the line $l_a$, and
$\gamma_b = \mathrm{sign}\,c$ for the line $l_b$. According to
(\ref{eq3.9}) and (\ref{eq3.10}), the real or imaginary parts of
$\lambda_\pm$ coincide at $l_{a,b}$ for negative or positive
$\gamma_{a,b}$, respectively.

\begin{figure}
\begin{center}
    \includegraphics[angle=0, width=0.8\textwidth]{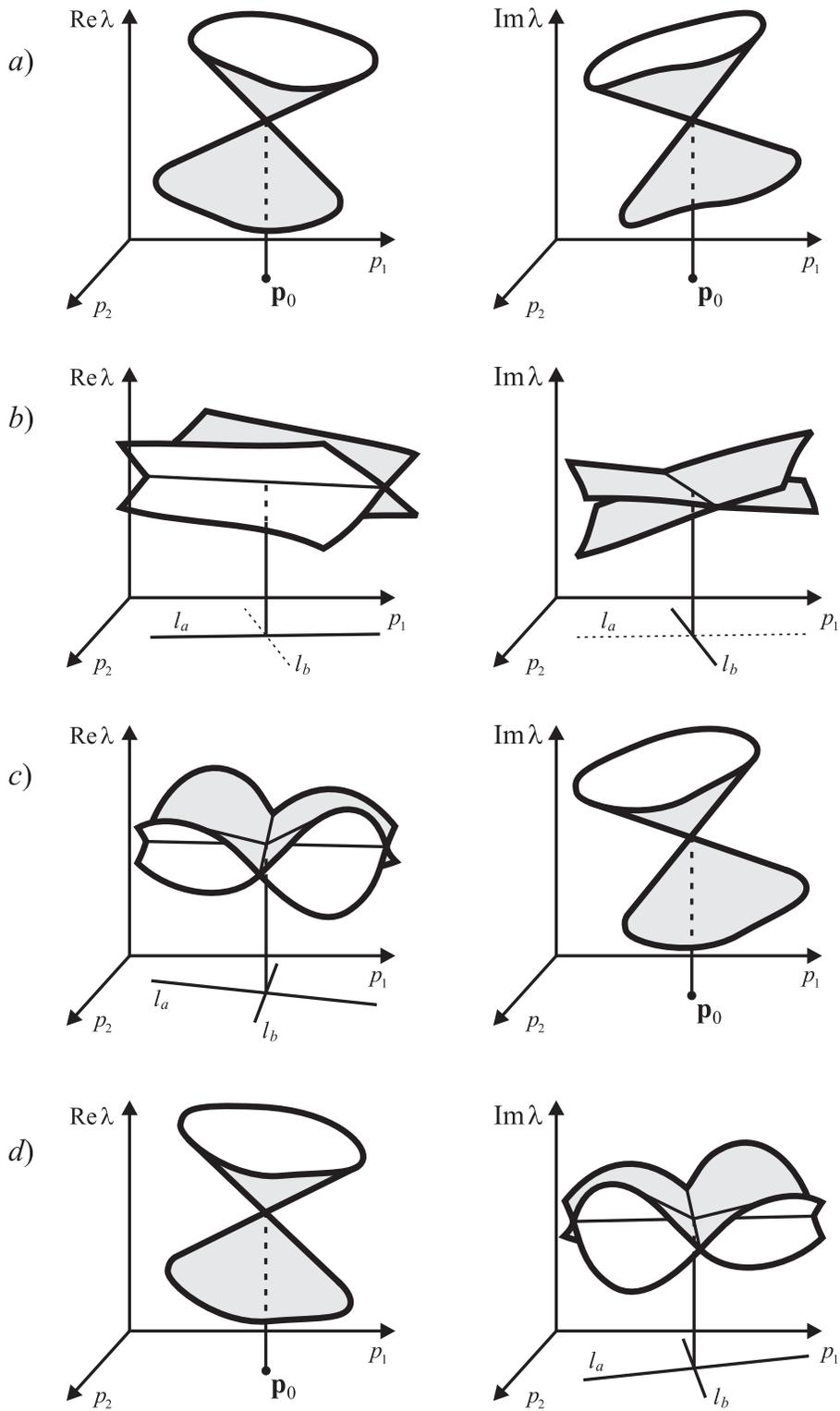}
\end{center}
\caption{Eigenvalue surfaces near a point of weak coupling.}
\label{fig3.2}
\end{figure}

One can distinguish four types of the graphs for
$\mathrm{Re}\,\lambda_\pm(p_1,p_2)$ and
$\mathrm{Im}\,\lambda_\pm(p_1,p_2)$ shown in Figure~\ref{fig3.2}.
In nondegenerate case $D' \ne 0$, the eigenvalues $\lambda_+$ and
$\lambda_-$ are different for all parameter values except the
initial point $p_{1,2} = p_{1,2}^0$. If $D' < 0$, the eigenvalue
surfaces are cones with non-elliptic cross-section, see
Figure~\ref{fig3.2}a. Other three types correspond to the case $D'
> 0$. If $\gamma_a < 0$ and $\gamma_b > 0$ then there is an
intersection of the real parts along the line $l_a$ and an
intersection of the imaginary parts along the line $l_b$ (in case
$\gamma_a > 0$ and $\gamma_b < 0$ the lines $l_a$ and $l_b$ are
interchanged), see Figure~\ref{fig3.2}b. If $\gamma_a < 0$ and
$\gamma_b < 0$ then the real parts intersect along the both lines
$l_a$ and $l_b$ forming a "cluster of shells", while there is no
intersections for the imaginary parts, see Figure~\ref{fig3.2}c.
Finally, if $\gamma_a > 0$ and $\gamma_b > 0$ then there is no
intersections for the real parts, while the imaginary parts
intersect along the both lines $l_a$ and $l_b$, see
Figure~\ref{fig3.2}d.

\section{Example}

As a physical example, we consider propagation of light in a
homogeneous non-magnetic crystal in the general case when the
crystal possesses natural optical activity (chirality) and
dichroism (absorption) in addition to biaxial birefringence, see
\cite{Berry2} for the general formulation. The optical properties
of the crystal are characterized by the inverse dielectric tensor
$\boldsymbol \eta$. The vectors of electric field $\bf E$ and
displacement $\bf D$ are related as \cite{LLP}
    \begin{equation}\label{eq2.19a}
    {\bf E}={\boldsymbol \eta}{\bf D}.
    \end{equation}
The tensor $\boldsymbol \eta$ is described by a non-Hermitian
complex matrix. The electric field $\bf E$ and magnetic field $\bf
H$ in the crystal are determined by Maxwell's equations \cite{LLP}
    \begin{equation}\label{eq2.19}
    {\rm rot} {\bf E}=-\frac{1}{c}\frac{\partial {\bf H}}{\partial t},
    ~~
    {\rm rot} {\bf H}=\frac{1}{c}\frac{\partial {\bf D}}{\partial
    t},
    \end{equation}
where $t$ is time and $c$ is the speed of light in vacuum.

A monochromatic plane wave of frequency $\omega$ that propagates
in a direction specified by a real unit vector ${\bf
s}=(s_1,s_2,s_3)$ has the form
    \begin{equation}\label{eq2.20}
    {\bf D}({\bf r},t)
    = {\bf D}({\bf s})\exp i\omega\!\left(\frac{n({\bf s})}{c}
    {\bf s}^T{\bf r}-t\right),\
    {\bf H}({\bf r},t)
    = {\bf H}({\bf s})\exp i\omega\!\left(\frac{n({\bf s})}{c}
    {\bf s}^T{\bf r}-t\right),
    \end{equation}
where $n({\bf s})$ is a refractive index, and ${\bf r} =
(x_1,x_2,x_3)$ is the real vector of spatial coordinates.
Substituting the wave (\ref{eq2.20}) into Maxwell's equations
(\ref{eq2.19}), we find
    \begin{equation}\label{eq2.21}
    {\bf H} = n [{\bf s}, {\boldsymbol \eta}{\bf D}], ~~
    {\bf D} = - n [{\bf s}, {\bf H}],
    \end{equation}
where square brackets indicate cross product of vectors
\cite{LLP}. With the vector $\bf H$ determined by the first
equation of (\ref{eq2.21}), the second equation of (\ref{eq2.21})
yields \cite{Berry2}
    \begin{equation}\label{eq2.22}
    -[{\bf s},[{\bf s}, {\boldsymbol \eta} {\bf D}({\bf s})]]={\boldsymbol
    \eta}{\bf D}({\bf s})-{\bf
    s}({\bf s}^T{\boldsymbol \eta}{\bf D}({\bf s}))=\frac{1}{n^2({\bf s})}{\bf D}({\bf s}).
    \end{equation}
Multiplying equation (\ref{eq2.22}) by the vector $\mathbf{s}^T$
from the left we find that for plane waves the vector ${\bf D}$ is
always orthogonal to the direction $\bf s$, i.e., ${\bf s}^T{\bf
D}({\bf s})=0$.

Since the quantity ${\bf s}^T{\boldsymbol \eta}{\bf D}({\bf s})$
is a scalar, we can write (\ref{eq2.22}) in the form of an
eigenvalue problem for the complex non-Hermitian matrix $\bf
A({\bf s})$ dependent on the vector of parameters ${\bf
s}=(s_1,s_2,s_3)$:
    \begin{equation}\label{eq2.24}
    {\bf A}{\bf u}=\lambda{\bf u},~~{\bf A}({\bf s})=({\bf I}-{\bf s}{\bf s}^T){\boldsymbol
    \eta}({\bf s}),
    \end{equation}
where $\lambda=n^{-2}$, ${\bf u}={\bf D}$, and ${\bf I}$ is the
identity matrix. Multiplying the matrix $\bf A$ by the vector $\bf
s$ from the left we conclude that ${\bf s}^T{\bf A}=0$, i.e., the
vector ${\bf s}$ is the left eigenvector with the eigenvalue
$\lambda=0$. Zero eigenvalue always exists, because $\det({\bf
I}-{\bf s}{\bf s}^T)\equiv 0$, if $\|{\bf s}\|=1$.

The matrix ${\bf A}({\bf s})$ defined by equation (\ref{eq2.24})
is a product of the matrix ${\bf I}-{\bf s}{\bf s}^T$ and the
inverse dielectric tensor ${\boldsymbol \eta}({\bf s})$. The
symmetric part of ${\boldsymbol \eta}$ constitutes the anisotropy
tensor describing the birefringence of the crystal. It is
represented by the complex symmetric matrix $\bf U$, which is
independent of the vector of parameters $\bf s$. The antisymmetric
part of ${\boldsymbol \eta}$ is determined by the optical activity
vector ${\bf g}({\bf s}) = (g_1,g_2,g_3)$, describing the
chirality (optical activity) of the crystal. It is represented by
the skew-symmetric matrix
\begin{equation}\label{eq2.25}
    {\bf G} = i\left(%
\begin{array}{ccc}
  0 & -g_3 & g_2 \\
  g_3 & 0 & -g_1 \\
  -g_2 & g_1 & 0 \\
\end{array}%
\right).
\end{equation}
The vector $\bf g$ is given by the expression ${\bf g}({\bf
s})={\boldsymbol \gamma}{\bf s}$, where $\boldsymbol \gamma$ is
the optical activity tensor represented by a symmetric complex
matrix. Thus, the matrix $\mathbf{G}(\mathbf{s})$ depends linearly
on the parameters $s_1$, $s_2$, $s_3$.

In the present formulation, the problem was studied analytically
and numerically in \cite{Berry2}. Below we present two specific
numerical examples in case of a non-diagonal matrix
$\boldsymbol\gamma$, for which the structure of singularities was
not fully investigated. Unlike \cite{Berry2}, where the reduction
to two dimensions was carried out, we work with the
three-dimensional form of problem (\ref{eq2.24}). Our intention
here is to give guidelines for using our theory by means of the
relatively simple $3\times 3$ matrix family, keeping in mind that
the main area of applications would be higher dimensional
problems.

As a first example, we choose the inverse dielectric tensor in the
form
    \begin{equation}\label{eq2.26}
    {\boldsymbol \eta} =
    \left(%
    \begin{array}{ccc}
    3 & 0 & 0\\
    0 & 1 & 0\\
    0 & 0 & 2\\
    \end{array}%
    \right)+
   i\left(%
    \begin{array}{ccc}
    0 & 1 & 2\\
    1 & 0 & 0\\
    2 & 0 & 0\\
    \end{array}%
    \right)+
   i\left(%
    \begin{array}{lll}
    0 & -s_1 & 0\\
    s_1 & 0 & -s_3\\
    0 & s_3 & 0\\
    \end{array}%
    \right)
    \end{equation}
where $s_3=\sqrt{1-s_1^2-s_2^2}$. The crystal defined by
(\ref{eq2.26}) is dichroic and optically active with the
non-diagonal matrix $\boldsymbol\gamma$. When $s_1=0$ and $s_2=0$
the spectrum of the matrix ${\bf A}$ consists of the double
eigenvalue $\lambda_0=2$ and the simple zero eigenvalue. The
double eigenvalue possesses the eigenvector ${\bf u}_0$ and
associated vector ${\bf u}_1$:
    \begin{equation}\label{eq2.27}
    {\bf u}_0=\left(%
    \begin{array}{c}
    i \\
    -1 \\
    0 \\
    \end{array}%
    \right),~~
    {\bf u}_1=\left(%
    \begin{array}{c}
    0 \\
    1 \\
    0 \\
    \end{array}%
    \right).
    \end{equation}
The eigenvector ${\bf v}_0$ and associated vector ${\bf v}_1$
corresponding to the double eigenvalue $\lambda_0=2$ of the
adjoint matrix ${\bf A}^*$ are
    \begin{equation}\label{eq2.28}
    {\bf v}_0=\left(%
    \begin{array}{c}
    i \\
    1 \\
    1+i/2 \\
    \end{array}%
    \right),~~
    {\bf v}_1=\left(%
    \begin{array}{c}
    i \\
    0 \\
    1/2-i/4 \\
    \end{array}%
    \right).
    \end{equation}
The vectors ${\bf u}_0$, ${\bf u}_1$ and ${\bf v}_0$, ${\bf v}_1$
satisfy the normalization and orthogonality conditions
(\ref{eq1.6}). Calculating the derivatives of the matrix ${\bf
A}(s_1,s_2)$ at the point ${\bf s}_0=(0,0,1)$ we obtain
    \begin{equation}\label{eq2.29}
    \frac{\partial {\bf A}}{\partial s_1}=\left(%
    \begin{array}{ccc}
    -2i & -2i & -2 \\
    i & 0 & 0 \\
    -3 & -i & -2i \\
    \end{array}%
    \right),~~
    \frac{\partial {\bf A}}{\partial s_2}=\left(%
    \begin{array}{ccc}
    0 & 0 & 0 \\
    -2i & -i & -2 \\
    -i & -1 & i \\
    \end{array}%
    \right).
    \end{equation}

    \begin{figure}
    \begin{center}
    \includegraphics[angle=0, width=1\textwidth]{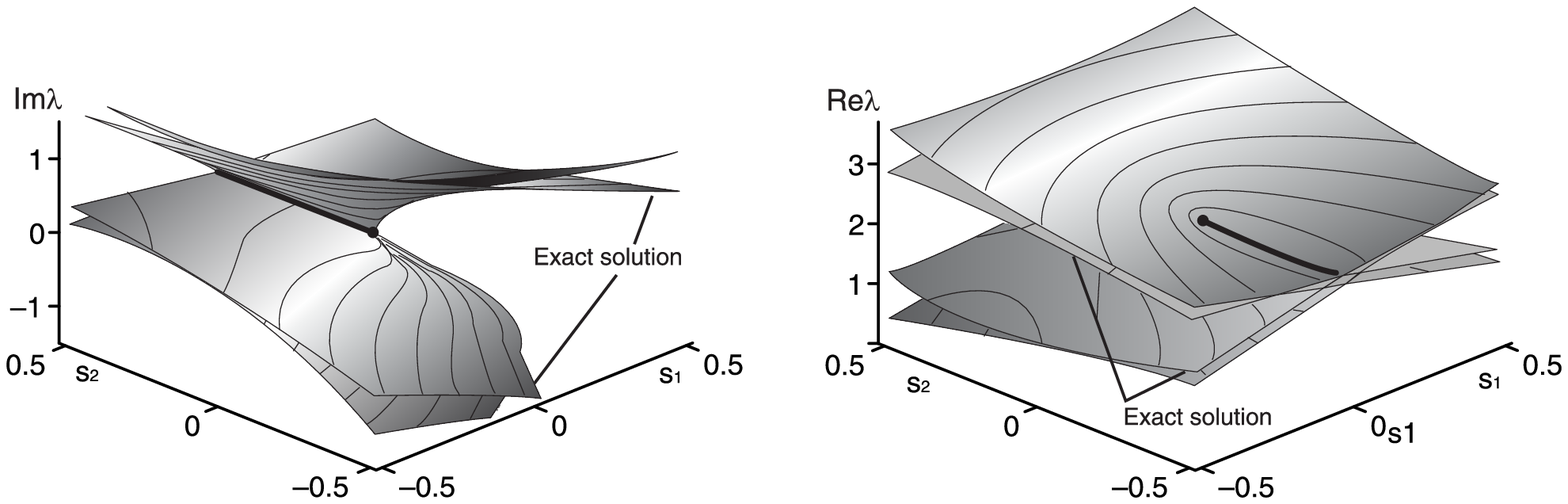}
    \end{center}
    \caption{Eigensurfaces of the crystal (\ref{eq2.26}) and their approximations.}
    \label{fig7}
    \end{figure}

Substitution of the derivatives (\ref{eq2.29}) together with the
vectors given by equations (\ref{eq2.27}) and (\ref{eq2.28}) into
the formulae (\ref{eq2.1}) and (\ref{eq2.1a}) yields the vectors
${\bf f}$, ${\bf g}$ and $\bf h$, $\bf r$ as
    \begin{equation}\label{eq2.30}
    {\bf f}=(0, 4),~~{\bf g}=(-4, 0),~~{\bf h}=(0,0),~~{\bf r}=(-4,0).
    \end{equation}
With the vectors (\ref{eq2.30}) we find from (\ref{eq2.8}) and
(\ref{eq2.9}) the approximations of the eigensurfaces ${\rm
Re}\lambda(s_1,s_2)$ and ${\rm Im}\lambda(s_1,s_2)$ in the
vicinity of the point ${\bf s}_0=(0,0,1)$:
    \begin{equation}\label{eq2.31}
    {\rm Re}\lambda_{\pm} = 2 \pm \sqrt{2s_2 +
    2\sqrt{s_1^2+s_2^2}},~~
    {\rm Im}\lambda_{\pm} =
    -2s_1\pm\sqrt{-2s_2+2\sqrt{s_1^2+s_2^2}}.
    \end{equation}
Calculation of the exact solution of the characteristic equation
for the matrix $\bf A$ with the inverse dielectric tensor
$\boldsymbol \eta$ defined by equation (\ref{eq2.26}) shows a good
agreement of the approximations (\ref{eq2.31}) with the numerical
solution, see Figure~{\ref{fig7}}. One can see that the both
surfaces of real and imaginary parts have a Whitney umbrella
singularity at the coupling point; the surfaces self-intersect
along different rays, which together constitute a straight line
when projected on parameter plane.

As a second numerical example, let us consider the inverse
dielectric tensor as
    \begin{equation}
    {\boldsymbol\eta}
    = \left(\begin{array}{ccc}
        1 & 0 & 1 \\
        0 & 1 & 0 \\
        1 & 0 & 4
    \end{array}\right)
    +i\left(\begin{array}{ccc}
        5 & 0 & 4 \\
        0 & 5 & 2 \\
        4 & 2 & 0
    \end{array}\right)
    +4i\left(\begin{array}{ccc}
        0 & -s_1-is_2 & is_3 \\
        s_1+is_2 & 0 & -s_3 \\
        -is_3 & s_3 & 0
    \end{array}\right).
    \label{ex2.1}
    \end{equation}
At $\mathbf{s} = (0,0,1)$, the matrix $\mathbf{A}$ has the double
eigenvalue $\lambda_0 = 1+5i$ with two eigenvectors and the simple
zero eigenvalue. The eigenvectors $\mathbf{u}_1$, $\mathbf{u}_2$
of $\lambda_0$ and the eigenvectors $\mathbf{v}_1$, $\mathbf{v}_2$
of $\overline{\lambda}_0$ for the adjoint matrix $\mathbf{A}^*$
are
    \begin{equation}
    \mathbf{u}_1
    = \left(\begin{array}{c}
        1 \\ 0 \\ 0
    \end{array}\right),\quad
    \mathbf{u}_2
    = \left(\begin{array}{c}
        0 \\ 1 \\ 0
    \end{array}\right),\quad
    \mathbf{v}_1
    = \left(\begin{array}{c}
        1 \\ 0 \\ \frac{-3-4i}{1-5i}
    \end{array}\right),\quad
    \mathbf{v}_2
    = \left(\begin{array}{c}
        0 \\ 1 \\ \frac{2i}{1-5i}
    \end{array}\right).
    \label{ex2.2}
    \end{equation}
These vectors satisfy normalization conditions (\ref{eq1.NC}).
Taking derivatives of the matrix $\mathbf{A}$ with respect to
parameters $s_1$ and $s_2$, where $s_3 = \sqrt{1-s_1^2-s_2^2}$,
and using formula (\ref{eq3.2}), we obtain
    \begin{equation}
    \mathbf{d}_{11} = (-2-8i, 0),\
    \mathbf{d}_{12} = (6i, -9-4i),\
    \mathbf{d}_{21} = (-10i, 7-4i),\
    \mathbf{d}_{22} = (0, -4i).
    \label{ex2.3}
    \end{equation}
Using (\ref{ex2.3}) in formulae (\ref{eq3.3})--(\ref{eq3.5}), we
find approximations for real and imaginary parts of two nonzero
eigenvalues $\lambda_\pm$ near the point $\mathbf{s} = (0,0,1)$ as
    \begin{equation}
    \mathrm{Re}\,\lambda_\pm
    = 1-s_1\pm
    \sqrt{(|c|+\mathrm{Re}\,c)/2},\ \
    \mathrm{Im}\,\lambda_\pm
    = 5-4s_1-2s_2\pm
    \sqrt{(|c|-\mathrm{Re}\,c)/2},
    \label{ex2.4}
    \end{equation}
where $c = (45+8i)s_1^2+128is_1s_2+(-83+8i)s_2^2$.

\begin{figure}
    \begin{center}
    \includegraphics[angle=0, width=1\textwidth]{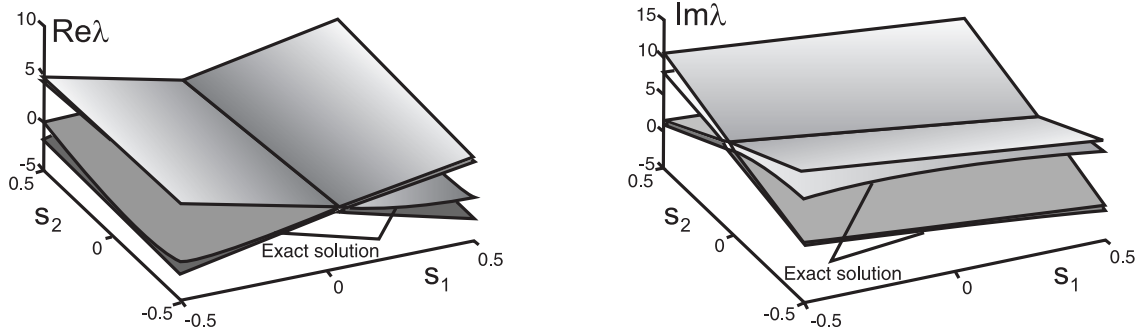}
    \end{center}
    \caption{Eigensurfaces of the crystal (\ref{ex2.1}) and their approximations.}
    \label{fig4.1}
    \end{figure}

Approximations of eigenvalue surfaces (\ref{ex2.4}) and the exact
solutions are presented in Figure~\ref{fig4.1}. The eigenvalue
surfaces have intersections both in
$(s_1,s_2,\mathrm{Re}\,\lambda)$ and
$(s_1,s_2,\mathrm{Im}\,\lambda)$ spaces. These intersections are
represented by two different lines $l_a$ and $l_b$ in parameter
space, see Figure~\ref{fig3.2}b.

\section{Conclusion}

A general theory of coupling of eigenvalues of complex matrices
smoothly depending on multiple real parameters has been presented.
Diabolic and exceptional points have been mathematically described
and general formulae for coupling of eigenvalues at these points
have been derived. This theory gives a clear and complete picture
of crossing and avoided crossing of eigenvalues with a change of
parameters. It has a very broad field of applications since any
physical system contains parameters. It is important that the
presented theory of coupling gives not only qualitative, but also
quantitative results on eigenvalue surfaces based only on the
information at the diabolic and exceptional points. This
information includes eigenvalues, eigenvectors and associated
vectors with derivatives of the system matrix taken at the
singular points. We emphasize that the developed methods provide a
firm basis for analysis of spectrum singularities of matrix
operators.

\section{Acknowledgement}

The work is supported by the research grants RFBR-NSFC
02-01-39004, RFBR 03-01-00161, CRDF-BRHE Y1-MP-06-19, and
CRDF-BRHE Y1-M-06-03.

\end{document}